\documentclass[aps,prb,twocolumn,superscriptaddress,
preprintnumbers,amsmath,amssymb
,floatfix,showpacs
]{revtex4}

\usepackage{graphicx}        
\usepackage{mathrsfs} 

\usepackage{graphicx}
\usepackage{dcolumn}
\usepackage{bm}
\usepackage{amsmath}
\usepackage{amssymb}
\usepackage{revsymb}

\begin{document}

\affiliation{
Department of Physics and Astronomy, Georgia State
University, Atlanta, Georgia 30303, USA}

\title{Spaser Action, Loss Compensation, and Stability in Plasmonic Systems with Gain}


\author{Mark I. Stockman}
\affiliation{
Department of Physics and Astronomy, Georgia State
University, Atlanta, Georgia 30303, USA}
\email{mstockman@gsu.edu}
\homepage{http://www.phy-astr.gsu.edu/stockman}

\date{\today}

\begin{abstract}
We demonstrate that the conditions of spaser generation and the full loss compensation in a resonant plasmonic-gain medium (metamaterial) are identical. Consequently, attempting the full compensation or overcompensation of losses by gain will lead to instability  and a transition to a spaser state. This will limit (clamp) the inversion and lead to the limitation on the maximum loss compensation achievable. The criterion of the loss overcompensation, leading to  the instability and spasing, is given in an analytical and universal (independent from system's geometry) form.
\end{abstract}


\maketitle

There is a tremendous interest  in nanoplasmonic systems with gain, which has been initiated by the introduction of the spaser \cite{Bergman_Stockman:2003_PRL_spaser}. Such systems consist of a metal nanoplasmonic component and a nanoscale gain medium (dye molecules, quantum dots, etc.) where the population inversion is created optically or electrically \cite{Bergman_Stockman:2003_PRL_spaser, Bergman_Stockman_Spaser_Patent_2009}. If the surface plasmon (SP) amplification by stimulated emission overcomes the loss, the  initial state of the system loses its stability, and a new, spasing state appears with a coherent SP population whose phase is established due to a  spontaneous symmetry breaking \cite{Stockman_JOPT_2010_Spaser_Nanoamplifier}. The spaser is a nanoscopic generator of coherent local optical fields  and their ultrafast nanoamplifier. 

There has been a very active development of the idea and principles of spaser. A nanoscopic spaser consisting of gold nanosphere core surrounded by a dielectric gain shell containing a laser dye has been demonstrated \cite{Noginov_et_al_Nature_2009_Spaser_Observation}. Surface plasmon polariton (SPP) spasers have been  demonstrated with one-dimensional \cite{Hill_et_al_Opt_Expr_2009_Polaritonic_Nanolaser} and  two-dimensional \cite{Oulton_Sorger_Zentgraf_Ma_Gladden_Dai_Bartal_Zhang_Nature_2009_Nanolaser} confinement. The pre-generation narrowing of the resonant line  in the lasing spaser has been observed \cite{Plum_Fedotov_Kuo_Tsai_Zheludev_Opt_Expr_2009_Toward_Lasing_Spaser}.

One of the most active research directions related to the spaser has been compensation of losses by gain in plasmonic waveguides and metamaterials, which is of principal importance due to high losses in the optical range of frequencies. Amplification of long-range SPPs in a gold strip waveguide in the proximity of a pumped dye solution has been demonstrated \cite{Leon_Berini_Nature_Photonics_2010_SPP_Amplification}. 
Amplified spontaneous emission of SPPs has been observed in gold nanofilms over an amplifying medium containing PbS quantum dots, where the reduction of the SPP propagation loss by up to 30 percent has taken place \cite{Zayats_et_al_OL35_1197_2010_SPP_Amplified_Spontaneous_Emission}. In a metamaterial consisting of split-ring resonators coupled to an optically-pumped InGaAs quantum well, a reduction of the transmission loss by $\approx 8$ percent has been observed \cite{Wegener_et_al_arXiv_2010_SRR_QW}. The full compensation and overcompensation of the optical transmission loss for a fishnet metamaterial containing a pumped dye dispersed in a polymer matrix has been observed \cite{Shalaev_et_al_Nature_2010_Loss_Free_Active_NIM}. This experiments has later been stated to be in agreement with a theory based on a Maxwell-Bloch equations  \cite{Hess_PRL_2010_Loss_Compensation}.

In this Letter we show that the full compensation or overcompensation of the optical loss loss in an active metamaterial (i.e., a nanostructured optical system of a finite size containing a gain medium) leads to an instability that is resolved by its spasing (i.e., due to becoming a spaser). We further show that the conditions of the complete resonant gain compensation (which is the only one explored either experimentally or theoretically so far) and the threshold condition of spasing are identical. This spasing limits (clamps)  the gain and, consequently, does not allow for the complete loss compensation (overcompensation) at any frequency. Additionally, this spasing in the gain metamaterial will show itself as enhanced (amplified) spontaneous emission and the coherent emission reminding the lasing spaser \cite{Zheludev_et_al_Nat_Phot_2008_Lasing_Spaser, Plum_Fedotov_Kuo_Tsai_Zheludev_Opt_Expr_2009_Toward_Lasing_Spaser}.

We will consider, for certainty, an isotropic and uniform metamaterial that, by definition, in a range of frequencies $\omega$ can be described by the effective permittivity $\bar\varepsilon(\omega)$ and permeability $\bar\mu(\omega)$. We will concentrate below on the loss compensation  for the optical electric responses; similar consideration with identical conclusions for the optical magnetic responses is straightforward. Consider a small piece of the metamaterial with sizes much greater that the unit cell but much smaller than the wavelength $\lambda$, which is a metamaterial itself. Let us subject this metamaterial to a uniform electric field $\mathbf E_0(\omega)$ oscillating with frequency $\omega$. We will denote the local field at a point $\mathbf r$ inside this metamaterial as $\mathbf e(\mathbf r, \omega)$.  For such a small piece of the metamaterial, a homogenization procedure gives an exact expression (see Ref.\ \onlinecite{Stockman_et_al_1999_Linear_and_Nonlinear_Maxwell_Garnett} and references cited therein)
\begin{equation}
\bar\varepsilon(\omega)=\frac{1}{V E_0(\omega)^2} \protect\int_V \varepsilon(\mathbf r, \omega) \left[\mathbf  e(\mathbf r,\omega)\right]^2 d^3 r~,
\label{homo}
\end{equation}
where $V$ is the volume of the metamaterial piece. 

Consider a frequency $\omega$ close to the resonance frequency $\omega_n$ of an $n$th plasmonic eigenmode. To be bright, this eigenmode must be dipolar.
Then the Green's function expansion \cite{Stockman:2002_PRL_control, Li_Stockman_PRB_2008_Time_Reversal} shows that the eigenmode's field  can be estimated as  as $\sim E_0 Q (R/L)^3\sim E_0 Q f$, where $L$ is the size of the metamaterial's elementary cell, and $R$ is the size and $f$ is the fill factor of the metal component. Realistically assuming that $Q f\gg 1$, we conclude that the resonant eigenmode's field $\mathbf  e(\mathbf r,\omega)\propto -\nabla\varphi_n(\mathbf r)$ dominates the local field. In this case, the effective permittivity (\ref{homo}) becomes
\begin{equation}
\bar\varepsilon(\omega)=\frac{\int_V \varepsilon(\mathbf r, \omega) \left[\mathbf  \nabla\varphi_n(\mathbf r)\right]^2 d^3 r}{\int_V  \left[\mathbf  \nabla\varphi_n(\mathbf r)\right]^2 d^3 r}
\label{homo1}
\end{equation}

The quasistatic eigenmode equation is \cite{Stockman:2001_PRL_Localization}
\begin{equation}
\nabla\theta(\mathbf r)\nabla\varphi_n(\mathbf r)=s_n\nabla^2\varphi_n(\mathbf r)~,
\label{eigen}
\end{equation}
where $s_n$ is the corresponding eigenvalue, and $\theta(\mathbf r)$ is the characteristic function that is equal to 1 inside the metal and 0 otherwise. The homogeneous Dirichlet-Neumann boundary conditions are implied. 

From Eq.\ (\ref{eigen}) one can easily find a relation
\begin{equation}
s_n=\frac{\int_V \theta(\mathbf r) \left|\nabla\varphi_n(\mathbf r) \right|^2 d^3 r} {\int_V \left|\nabla\varphi_n(\mathbf r) \right|^2 d^3 r}~,
\label{norm}
\end{equation}
from which it follows, in particular, that $1\ge s_n\ge 0$. The resonant frequency, $\omega=\omega_n$, is defined by the equality 
\begin{equation}
s_n=\mathrm{Re}\,s(\omega)~,~~~ s(\omega)\equiv\frac{\varepsilon_h(\omega)} {\varepsilon_h(\omega)-\varepsilon_m(\omega)}~,  
\label{s_n}
\end{equation}
where $s(\omega)$ is Bergman's spectral parameter,  $\varepsilon_m(\omega)$ is the permittivity of the metal, and $\varepsilon_h(\omega)$ is that of the surrounding host with the gain chromophore centers. 

In the case of the full inversion (maximum gain) and in the exact resonance, the host medium permittivity acquires the imaginary part responsible for the stimulated emission as given by the standard expression
\begin{equation}
\varepsilon_h(\omega)=\varepsilon_d-i \frac{4\pi}{3}\frac{\left|\mathbf d_{12}\right|^2 n_c}{\hbar \Gamma_{12}}~,
\label{host}
\end{equation}
where $\varepsilon_d=\mathrm{Re}\,\varepsilon_h$,   $\mathbf d_{12}$ is a dipole matrix element of the gain transition in a chromophore center of the gain medium,  $\Gamma_{12}$ is a spectral width of this transition, and $n_c$ is the concentration of these centers. 

Using Eqs.\ (\ref{homo1}) and (\ref{norm}), it is straightforward to show that the effective permittivity (\ref{homo1}) simplifies exactly to
\begin{equation}
\bar\varepsilon(\omega)=s_n \varepsilon_m(\omega) + (1-s_n) \varepsilon_h(\omega)~.
\label{homo2}
\end{equation}
The condition for the full electric loss compensation in the metamaterial and amplification (overcompensation) at the resonant frequency $\omega=\omega_n$  is
\begin{equation}
 \mathrm{Im}\,\bar\varepsilon(\omega)=s_n\mathrm{Im}\,\varepsilon_m(\omega) - \frac{4\pi}{3}\frac{\left|\mathbf d_{12}\right|^2 n_c (1-s_n)}{\hbar \Gamma_{12}} \le 0~.
\label{amplification}
\end{equation}
Finally,  taking into account Eq.\ (\ref{s_n}) and that $s_n>0$ and $\mathrm{Im}\,\varepsilon_m(\omega)>0$, we obtain from Eq.\ (\ref{amplification}) the condition of the loss (over)compensation as
\begin{equation}
\frac{4\pi}{3}\frac{\left|\mathbf d_{12}\right|^2 n_c\left[1-\mathrm{Re}\,s(\omega)\right]}{\hbar \Gamma_{12}\mathrm{Re}\,s(\omega)\mathrm{Im}\,\varepsilon_m(\omega)}\ge 1~,
\label{result1}
\end{equation}
where the strict inequality corresponds to the overcompensation and net amplification. In Eq.\ (\ref{host}) we have assumed non-polarized gain transitions. If these transitions are all polarized along the excitation electric field, the concentration $n_c$ should be multiplied by a factor of $3$.

This is a fundamental condition, which is precise (for the problem under consideration) and general. Nevertheless, it is fully analytical and, actually, very simple. Remarkably, it depends only on the material characteristics and does not contain any geometric properties of the metamaterial system or the local fields. In particular, the hot spots, which are prominent in the local fields of nanostructures \cite{Stockman_et_al_PRB_1996_Inhomogeneous_Localization_Hot_Spots, Stockman:2001_PRL_Localization}, are completely averaged out due to the integrations in Eqs.\ (\ref{homo}) and (\ref{homo1}). We note that this implies that taking into account the gain enhancement due to the local field effects in Ref.\ \onlinecite{Hess_PRL_2010_Loss_Compensation} is erroneous. 

The condition (\ref{result1}) is completely non-relativistic (quasistatic) -- it does not contain speed of light $c$, which is characteristic of the spaser. It is useful the express this condition also in terms of the total extinction cross section $\sigma_e(\omega)$ (where $\omega$ is the central resonance frequency)  of a chromophore of the gain medium as
\begin{equation}
\frac{c \sigma(\omega) \sqrt{\varepsilon_d} n_c \left[1-\mathrm{Re}\,s(\omega)\right]} {\omega \mathrm{Re}\,s(\omega)\mathrm{Im}\,\varepsilon_m(\omega)} \ge 1~.
\label{sigma}
\end{equation}

It is of fundamental importance to compare this condition of the full loss  (over)compensation with the spasing condition \cite{Bergman_Stockman:2003_PRL_spaser}. This criterion of spasing, which we will use in the form of Eq.\ (14) of Ref.\  \onlinecite{Stockman_JOPT_2010_Spaser_Nanoamplifier}, is fully applicable for the considered metamaterial. For the zero detuning between the gain medium and the SP eigenmode, this criterion can be exactly expressed as \cite{Stockman_JOPT_2010_Spaser_Nanoamplifier}
\begin{equation}
\frac{4\pi}{3}\frac{\left|\mathbf{d}_{12}\right|^2 \mathrm{Re}\, s(\omega)} {\hbar \gamma_n \Gamma_{12} \mathrm{Re}\, s^\prime(\omega) } \int_V \left|\nabla \varphi_n(\mathbf{r})\right|^2\rho(\mathbf{r})d^3r \ge 1~
\label{spasing}
\end{equation}
where $\gamma_n=\left.\mathrm{Im}\, s(\omega)\right/\mathrm{Re}\, s^\prime(\omega)$ is the decay rate of the SPs at a frequency $\omega$, $ s^\prime(\omega)\equiv\partial s(\omega)/\partial\omega$, and $\rho(\mathbf{r})$ is the density of the gain medium chromophores. 

The field quantization in general and SP-field quantization in particular can only be carried out consistently when the energy loss is small enough \cite{Bergman_Stockman:2003_PRL_spaser}. In our case, this implies that the quality factor $Q=-\left.\mathrm{Re}\,\varepsilon_m\right/ \mathrm{Im}\,\varepsilon_m\gg 1$. Otherwise the field energy needed for the quantization is not conserved and, actually, cannot be introduced \cite{Landau_Lifshitz_Electrodynamics_Continuous:1984}. For $Q\gg1$, we have, with a good accuracy, 
\begin{equation}
\gamma_n=\frac{\mathrm{Im}\,\varepsilon_m(\omega)}{\mathrm{Re}\,\varepsilon^\prime_m(\omega)}~,~~~ \mathrm{Re}\, s^\prime(\omega)=\frac{1}{\varepsilon_d} \left[\mathrm{Re}\, s(\omega)\right]^2 \mathrm{Re}\,
\varepsilon^\prime_m(\omega)~,
\label{simplified}
\end{equation}
where $\varepsilon^\prime_m(\omega)=\left.\partial \varepsilon_m(\omega)\right/ \partial\omega$. Substituting this into Eq.\ (\ref{spasing}), we obtain for the spasing condition
\begin{equation}
\frac{4\pi}{3}\frac{\left|\mathbf{d}_{12}\right|^2} {\hbar \Gamma_{12} \mathrm{Re}\, s(\omega) \mathrm{Im}\,\varepsilon_m(\omega) } \int_V \left|\nabla \varphi_n(\mathbf{r})\right|^2\rho(\mathbf{r})d^3r \ge 1~.
\label{criterion}
\end{equation}

The modal field in Eq.\ (\ref{criterion}) is normalized as $\int_V \left|\nabla \varphi_n(\mathbf{r})\right|^2 d^3r = 1$.  Taking this and Eq.\ (\ref{norm}) into account and assuming that $ \rho_n(\mathbf{r})=\left[1-\theta(\mathbf r)\right]n_c$, i.e.,  the chromophores are distributed in the dielectric with a constant density $n_c$, we {\em exactly} reduce Eq.\ (\ref{criterion}) to the form of Eq.\ (\ref{result1}). This brings us to an important conclusion: {\em the full compensation (overcompensation) of the optical losses in a metamaterial and the spasing occur under precisely the same conditions}.  Inequality (\ref{result1}) is the criterion for both the loss (over)compensation and spasing. 

This fact of the equivalence of the full loss compensation and spasing is intimately related to the general criteria of the the thermodynamic stability with respect to small fluctuations  of electric and magnetic fields --   see Chap. IX of  Ref.\ \onlinecite{Landau_Lifshitz_Electrodynamics_Continuous:1984}) 
\begin{equation}
\mathrm{Im}\,\bar\varepsilon(\omega)>0~,~~~\mathrm{Im}\,\bar\mu(\omega)>0~,
\label{stability}
\end{equation}
which must be {\em strict} inequalities for all frequencies. 

The first of these conditions is opposite to Eq.\ (\ref{result1}). This has a transparent meaning: the electrical instability of the system is resolved by its spasing that limits (clamps) the gain and population inversion  making {\em the net gain to be precisely zero} \cite{Stockman_JOPT_2010_Spaser_Nanoamplifier}. This makes the complete loss compensation and its overcompensation impossible in a metamaterial with a feedback, which is created by the facets of the system and its internal inhomogeneities. 

\begin{figure}
\centering
\includegraphics[width=.3\textwidth]
{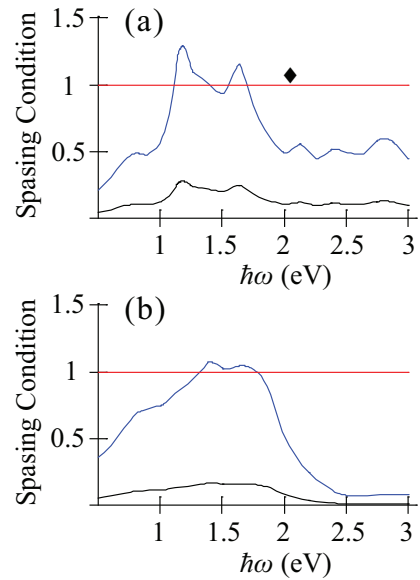}
\caption{\label{Spasing_Criterion}
Spasing criterion as a function of optical frequency $\omega$. The straight line (red on line) represents the threshold for the spasing and full loss compensation, which take place for the curve segments above it. (a) Computations for silver.  The chromophore concentration is $n_c=6\times 10^{18}~\mathrm{cm^{-3}}$ for the lower curve (black) and  $n_c=3.9\times 10^{19}~\mathrm{cm^{-3}}$ for the upper curve (blue on line). The black diamond shows the value of the spasing criterion for the conditions of Ref.\ \onlinecite{Noginov_et_al_PRL_2008_SPP_Stimulated_Emision} -- see the text.
(b) Computations for gold. The chromophore concentration is $n_c=3\times 10^{19}~\mathrm{cm^{-3}}$ for the lower curve (black) and  $n_c=2\times 10^{20}~\mathrm{cm^{-3}}$ for the upper curve (blue on line).
}
\end{figure}

Because the loss (over)compensation condition (\ref{result1}), which is also the spasing condition, is geometry-independent, it is useful to illustrate it for commonly used plasmonic metals, gold and silver \cite{Johnson:1972_Silver}. For the gain medium chromophores, we will use a reasonable set of  parameters, which we will, for the sake of comparison,  adapt from Ref.\ \onlinecite{Hess_PRL_2010_Loss_Compensation}: $\Gamma_{12}=5\times 10^{13}~\mathrm{s^{-1}}$ and $d_{12}=4.3\times 10^{-18}$ esu. The results of computations are shown in Fig.\ \ref{Spasing_Criterion}. For silver as a metal and $n_c=6\times 10^{18}~\mathrm{cm^{-3}}$, the corresponding lower (black) curve in panel (a) does not reach the value of $1$, implying that no full loss compensation is achieved. In contrast, for a higher but still very realistic concentration of $n_c=3.9\times 10^{19}~\mathrm{cm^{-3}}$,  the upper curve in  Fig.\ \ref{Spasing_Criterion} (a) does cross the threshold line in the near-infrared region. Above the threshold area, there will be the instability and the onset of the spasing. As  Fig.\ \ref{Spasing_Criterion} (b) demonstrates, for gold the spasing is at higher, but still realistic, chromophore concentrations.

Now let us discuss the implications of our results for the research published recently on the gain metamaterials. We start with the recent theoretical paper \cite{Hess_PRL_2010_Loss_Compensation}  that summarizes ``... We show that appropriate placing of optically pumped laser dyes (gain) into the metamaterial structure results in a frequency band where the nonbianisotropic metamaterial becomes amplifying. In that region both the real and the imaginary part of the effective refractive index become simultaneously negative and the figure of merit diverges at two distinct frequency points.'' 

In light of the present results, this conclusion of Ref.\ \onlinecite{Hess_PRL_2010_Loss_Compensation} is incorrect.  In reality, such a regime brings about the instability and spasing for the region of negative loss, $\mathrm{Im}\,\bar\varepsilon\le 0$, which is interpreted in Ref.\ \onlinecite{Hess_PRL_2010_Loss_Compensation} as the loss compensation. The full quantum mechanical theory  \cite{Stockman_JOPT_2010_Spaser_Nanoamplifier} of the spasing in gain nanoplasmonic systems shows that within the band of the spasing or (over)compensation, the population inversion is not defined by pumping and cannot be arbitrarily large. It is determined self-consistently by the processes of the stimulated emission of SPs and their relaxation. The stationary spasing decreases the inversion until it eliminates the net gain completely \cite{Stockman_JOPT_2010_Spaser_Nanoamplifier}. Theory of Ref.\ \onlinecite{Hess_PRL_2010_Loss_Compensation} misses an equation for the coherent SP field and, therefore, fails to describe the onset of spasing, which is a non-equilibrium second-order phase transition.

To carry out a quantitative comparison with Ref.\ \onlinecite{Hess_PRL_2010_Loss_Compensation}, we turn to Fig.\ \ref{Spasing_Criterion} (a) where the lower (black) curve corresponds to the nominal value of $n_c=6\times 10^{18}~\mathrm{cm^{-3}}$ used in Ref.\ \onlinecite{Hess_PRL_2010_Loss_Compensation}. There is no full loss compensation and spasing, which is explained by the fact that  Ref.\ \onlinecite{Hess_PRL_2010_Loss_Compensation} uses, as a close inspection shows, the gain dipoles parallel to the field and the local field enhancement [the latter, actually, is eliminated by the space integration -- see our discussion after Eq.\ (\ref{result1})]. This is equivalent to increasing in our formulas the concentration of the chromophores to  $n_c=3.9\times 10^{19}~\mathrm{cm^{-3}}$, which corresponds to the upper curve in  Fig.\ \ref{Spasing_Criterion} (a). This curve rises above the threshold line exactly in the same (infra)red region as in Ref.\ \onlinecite{Hess_PRL_2010_Loss_Compensation}. In reality, above the threshold there will be spasing causing the zero net gain and not a loss compensation. 

The complete loss compensation is also claimed in the recent experimental paper \cite{Shalaev_et_al_Nature_2010_Loss_Free_Active_NIM} where the system was actually a nanofilm and not a three-dimensional metamaterial. Consequently, the effective index parameters have been found from the transmission data by comparison to theory. For the chromophore concentrations used, it is, in principle, possible that the threshold of the overcompensation and spasing has been exceeded. Because in such a nanostructure the local fields are confined near the metal, the spaser feedback is all but unavoidable. In this case, the system should spase, which will cause the reduction (clamping) of inversion and the loss of gain. 

In an experimental study of the lasing spaser \cite{Plum_Fedotov_Kuo_Tsai_Zheludev_Opt_Expr_2009_Toward_Lasing_Spaser}, a nanofilm of PbS quantum dots (QDs) was positioned over a two-dimensional metamaterial consisting of an array of negative split ring resonators. When the QDs were optically pumped, the system exhibited an increase of the transmitted light intensity on the background of a strong luminescence of the QDs but apparently did not reach
the lasing threshold. The polarization dependent loss compensation was only $\sim 1$ \%. Similarly, for an array of split ring resonators over a resonant quantum well, where the inverted electron-hole population was  excited optically \cite{Wegener_et_al_arXiv_2010_SRR_QW}, the loss compensation did not exceed $\sim 8$ \%. The relatively low loss compensation in these papers may be due either to random spasing and/or spontaneous or amplified spontaneous emission enhanced by this plasmonic array, which reduces the population inversion.

A dramatic example of possible random spasing is presented in Ref.\ \onlinecite{Noginov_et_al_PRL_2008_SPP_Stimulated_Emision}. The  system studied was a Kretschmann-geometry setup \cite{Kretschmann_Raether_Z_Naturforsch_1968_SPP_Coupling_to_Light} with an added $\sim 1 \mathrm{\mu m}$ polymer film containing Rodamine 6G dye in the  $n_c=1.2\times 10^{19}~\mathrm{cm^{-3}} $ concentration. When the dye was pumped, there was outcoupling of radiation in a  range of angles. This was a threshold phenomenon with the threshold increasing with the Kretschmann angle.  At the maximum of the pumping intensity, the widest range of the outcoupling angles was observed, and the frequency spectrum at every angle narrowed to a peak near a single frequency $\hbar\omega\approx 2.1$ eV. This can be explained by the spasing where the feedback is provided by roughness of the metal. At the high pumping, the localized SPs with the highest threshold start to spase near a single frequency.  Because of their sub-wavelength size, the Kretschmann phase-matching condition is relaxed, and the radiation is outcoupled into a wide range of angles. Substituting the above-given parameters of the dye and the extinction cross section $\sigma_e=4\times 10^{-16}~\mathrm{cm^2}$ into Eq.\ (\ref{sigma}), we obtain a point shown by the black diamond in Fig.\ \ref{Spasing_Criterion}, which is clearly above the threshold, supporting our assertion of the spasing. Likewise, the amplified spontaneous emission and, possibly spasing, appear to have prevented the full loss compensation in a SPP system of Ref.\ \onlinecite{Zayats_et_al_OL35_1197_2010_SPP_Amplified_Spontaneous_Emission}. Note that if the feedback is thoroughly reduced as in the long-range SPP waveguide \cite{Leon_Berini_Nature_Photonics_2010_SPP_Amplification}, then the amplification instead of the spasing can be achieved.

Concluding, we have fundamentally established that the conditions of the full loss compensation (overcompensation) and spasing in plasmonic metamaterials are identical. These conditions are analytical and universal, i.e.,  independent from the metamaterial geometry. Due to the feedback inherent in the inhomogeneous metamaterials and waveguides, this implies that an attempt of the full loss compensation (over-compensation) in actuality brings about spasing that eliminates the net gain and precludes the full loss compensation. 

I am grateful to M.\ Noginov and N.\ Noginova for valuable discussions. This work was supported by a grant from the Chemical Sciences, Biosciences and Geosciences Division of the Office of Basic Energy Sciences, Office of Science, U.S. Department of Energy, and a grant from US-Israel Binational Science Foundation.


\end{document}